\documentstyle[aps,prl,epsf,twocolumn]{revtex}
\begin{document}
\draft
\preprint{draft}

\long\def\wideabs#1{\twocolumn[\hsize\textwidth\columnwidth\hsize%
\csname @twocolumnfalse\endcsname #1 \vskip1pc]}

\renewcommand{\dbltopfraction}{.99}
\renewcommand{\dblfloatpagefraction}{.99}
\renewcommand{\textfraction}{.01}

\title{Continuum-type stability balloon in oscillated granular layers}

\author{John R. de Bruyn\cite{perm}, C. Bizon, M. D. Shattuck, D. Goldman,
J. B. Swift, and Harry L. Swinney } \address{Center for Nonlinear
Dynamics and Department of Physics, University of Texas, Austin TX
78712} 
\date{\today} 
\wideabs{
\maketitle
\begin{abstract}

The stability of convection rolls in a fluid heated from below is
limited by secondary instabilities, including the skew-varicose and
crossroll instabilities. We observe a stability boundary defined by
the same instabilities in stripe patterns in a vertically oscillated
granular layer. Molecular dynamics simulations show that the mechanism
of the skew-varicose instability in granular patterns is similar to
that in convection. These results suggest that pattern formation in
granular media can be described by continuum models analogous to those
used in fluid systems.

\end{abstract}

}

A system comprised of a large number of discrete grains behaves like a
fluid under some conditions and like a solid under others, but can
also display behavior unique to the granular state \cite{jnb96}.
Despite substantial recent interest in both the statics and dynamics
of granular systems, there is no unifying theoretical description of
granular materials, and it has been argued that a general local,
continuum description of granular media, analogous to hydrodynamics,
is unlikely to exist \cite{jnb96,dlk95,haff83}. In this Letter we
report on a study of the stability of stripe patterns that form when a
horizontal layer of granular material is oscillated vertically
\cite{mus95,ums97,ums96}.  In both experiments and simulations, we
observe behavior strikingly similar to that seen in fluid dynamical
systems \cite{ch93}. Our results suggest that a continuum description
of pattern formation in granular media is possible.

Recent experiments on pattern formation in oscillated granular layers
\cite{mus95,ums97,ums96} have demonstrated the existence of stripe,
square and hexagonal patterns \cite{mus95,ums97}, as well as localized
structures called oscillons \cite{ums96}, as the frequency and
amplitude of the vibration and the layer thickness are varied. At
onset, the patterns are subharmonic, that is, they oscillate with a
frequency equal to half the driving frequency. The patterns are
similar in spatial structure to those observed in fluid dynamical
systems, most notably parallel convection rolls in a thin layer of
fluid heated from below (Rayleigh-B\'enard convection)
\cite{ch93,busse} and standing surface waves in a vertically
oscillated liquid layer (the Faraday instability) \cite{ch93,kg96}.
Molecular dynamics simulations \cite{lcrd96,bssms98} have reproduced
the various granular patterns at values of the control parameters
equal to those used in the experiments \cite{bssms98}. Some aspects of
the granular patterns have been described using various models for the
dynamics \cite{shin97}, amplitude equations \cite{ta97}, and iterative
maps \cite{vo97}.  However, a rigorously derived theoretical
description of pattern formation in oscillated granular media does not
yet exist, and the extent to which the well-developed understanding of
patterns in fluid systems \cite{ch93} can be applied to granular
material has not been established.

Our experimental apparatus is similar to that used in Refs.
\cite{mus95,ums97}.  Bronze spheres 150-180 $\mu$m in diameter are
contained in a cylindrical cell 14.7 cm in diameter. The cell sidewall
and top lid are made of Plexiglas. The aluminum bottom plate is flat
out to a radius of 5.0 cm, then slopes upwards to the sidewall at an
angle of 2$^\circ$.  As do fluid convection rolls, the stripes prefer
to form perpendicular to the cell wall; the ``beach'' reduces the
effect of the sidewall on their orientation. The cell is evacuated,
and an electromagnetic shaker, driven sinusoidally, oscillates the
cell vertically at a frequency $f$.  $\Gamma$, the amplitude of the
acceleration relative to the gravitational acceleration $g$, is
determined using an accelerometer mounted on the bottom of the
cell. The pattern is illuminated with a ring of LEDs encircling the
cell and strobed at $f/2$.  Images of the patterns are recorded using
a digital video camera mounted above the cell.

Experiments were performed on layers of thickness $N$ (scaled by the
mean particle diameter) in the range $6.6 < N < 11.3$ , in a frequency
range over which stripe patterns formed. The layer was flat for
$\Gamma < \Gamma_c \approx 2.5 $, at which point stripes appeared via
a subcritical bifurcation. Typically, $\Gamma$ was increased suddenly
from below onset to a value in the stripe regime. We observed
time-dependent patterns which were predominantly stripe-like but with
curvature, and which contained point defects, grain boundaries, spiral
defects\cite{ums97}, and sidewall foci. Stripes persisted as $\Gamma$
was increased, until the system (for most conditions) underwent a
subcritical transition to hexagons \cite{mus95}. Below this transition
we observed two secondary instabilities of the stripes --- the
skew-varicose and crossroll instabilities \cite{ch93,busse} --- which
limited the wavenumber range over which stripes were stable at a given
frequency.  Similar phenomena occured in cells without a beach.

Figure \ref{sv} shows the localized skew-varicose instability observed
when the local wave number of the pattern becomes too large. An
initially straight pattern of stripes develops a distortion.
Eventually one or more stripes pinch off and dislocation defects form,
which then propagate away by a combination of climbing and
gliding. Since one or more stripes are destroyed by this process, it
leads to a decrease in the local wave number.  Analogous
behavior has been observed in Rayleigh-B\'enard convection
\cite{pcl85,svrbc} and the Faraday instability \cite{ef94}.

\begin{figure*}[htbp] 
\epsfxsize=3.in 
\centerline{\epsffile{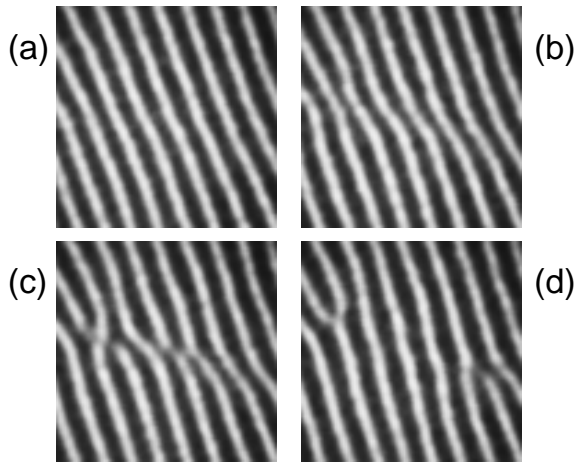}}
\vskip 0.1in 
\caption{A sequence of images showing the skew-varicose instability of
a stripe pattern and the formation of a pair of defects.  (a) Time
$t=-4.0$ s, straight stripes with wave number $\left<k\right>= 10.8$
cm$^{-1}$; (b) $t=0$ s, skew-varicose distortion is visible; (c)
$t=0.5$ s, defects form; (d) $t=1.0$ s, defects move away, leaving a
stripe pattern with $\left<k\right>= 9.9 $ cm$^{-1}$. $\Gamma = 3.10$,
$f = 35$ Hz, $N = 9.2$, and a 4.7 cm square region of the cell is
shown.}
\label{sv} 
\end{figure*}

\begin{figure*}[htbp]
\epsfxsize=3.in
\centerline{\epsffile{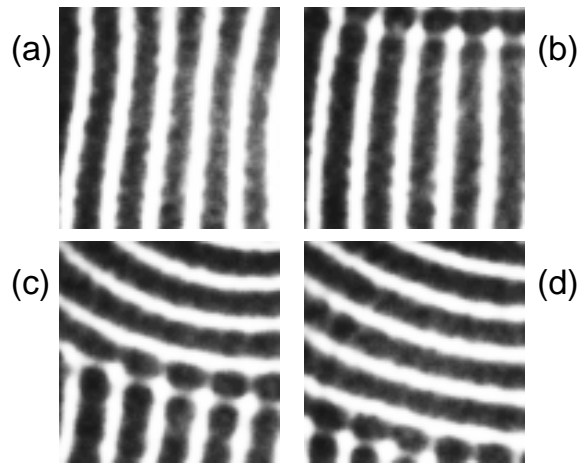}}
\vskip 0.1in
\caption{A sequence of images showing the crossroll instability of a
stripe pattern. (a) Time $t=-2.4$ s, straight stripes; (b) $t=0$ s,
crossrolls invade from the top edge of the image; in the center of the
image $\left<k\right> = 6.8$ cm$^{-1}$; (c) $t=3.5$ s; (d) $t=5.0$ s,
the pattern is predominantly perpendicular stripes with $\left<
k\right> = 7.0$ cm$^{-1}$.  $\Gamma = 2.82$, $f = 26.8$ Hz, $N =
11.3$, and the region shown is 4.7 cm square. }
\label{cr}
\end{figure*}

Figure \ref{cr} shows an example of a crossroll instability, which is
again analogous to that seen in fluid convection
\cite{pcl85,crrbc}. Crossrolls are observed to occur via two
processes, and, for most conditions, occur when the local wave number
becomes too small.  Squares and/or perpendicular stripes can develop
locally, or, as in Fig. \ref{cr}, a region of stripes can be invaded
by a region of perpendicularly oriented stripes.

\begin{figure*}[htbp]
\epsfxsize=3.375in
\centerline{\epsffile{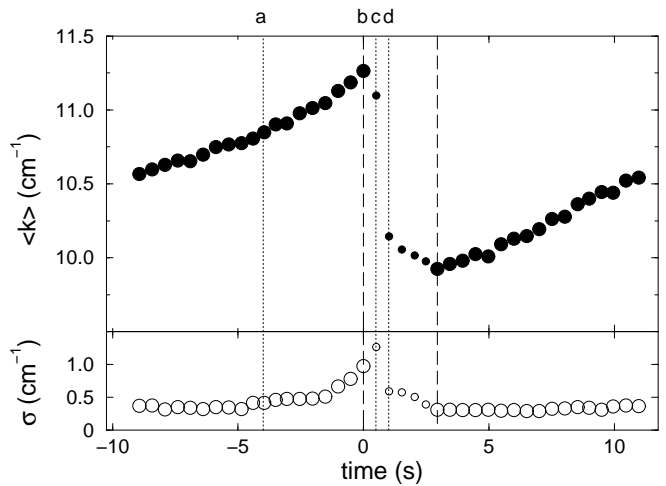}}
\vskip 0.1in
\caption{The average wave number $\left<k\right>$ (top) and standard
deviation $\sigma$ (bottom) for the skew-varicose event illustrated in
Fig. \protect\ref{sv}.  The times corresponding to the four images
shown in Fig. \protect\ref{sv} are indicated above the graph. Between
the dashed lines the pattern is not locally stripe-like due to the
presence of defects, and the values of $\left<k\right>$ and $\sigma$
are plotted as small symbols only to illustrate the change that occurs
when the defects form.}
\label{ktsv}
\end{figure*}

To determine the wave number $k$ at which the stripe pattern becomes
unstable to these instabilities, we used the method of
Ref. \cite{emb97} to determine $k$ at each point in a small region of
the pattern in which the instability occurs.  $k$ was averaged over
the region of interest for a time series of images bracketing the
instability. The mean wave number $\left< k \right>$ is plotted in
Fig. \ref{ktsv} for the skew-varicose event of Fig. \ref{sv}. $\left<
k \right>$ increases up to the onset of the instability, then suddenly
changes when a stripe is pinched off and defects form. The standard
deviation $\sigma$ of $\left< k \right>$ increases as the pattern
gradually becomes distorted and is relatively large immediately
following the instability.  After the defects have traveled away, a
stripe pattern with a smaller wave number remains and $\sigma$ is
again low. In Fourier-space the skew-varicose instability is signalled
by the local appearance of Fourier power at an oblique angle ---
typically 15$^\circ$ to 25$^\circ$ --- to the wave vector of the
original stripe pattern.

\begin{figure*}[htbp]
\epsfxsize=3.375in
\centerline{\epsffile{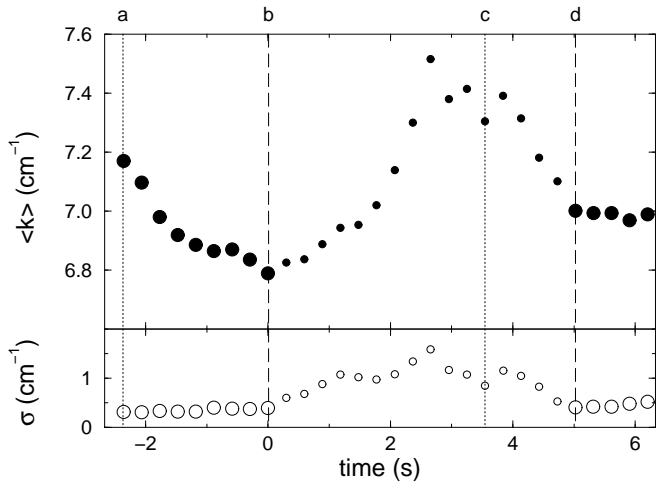}}
\vskip 0.1in
\caption{The average wave number $\left<k\right>$ (top) and standard
deviation $\sigma$ (bottom) for the crossroll event of
Fig. \protect\ref{cr}.  The times corresponding to the images shown in
Fig.  \protect\ref{cr} are indicated above the graph. Values of
$\left<k\right>$ and $\sigma$ between the dashed lines are plotted as
small symbols to illustrate the qualitative change that occurs when
the crossrolls invade.}
\label{ktcr}
\end{figure*}

Figure \ref{ktcr} shows analogous data for the crossroll event of
Fig. \ref{cr}. In this case $\left< k \right>$ decreases prior to the
onset of the instability. When the crossrolls invade, the local
pattern is disrupted and $\sigma$ increases. At the end of the time
interval shown, the region contains a pattern of straight stripes
perpendicular to and with $k$ larger than the original stripes.  In
Fourier space, the development of crossrolls is accompanied by the
appearance of power at 90$^\circ$ to the original wave vector.

By analyzing a number of similar events over a range of $\Gamma$, $f$,
and $N$, we can construct a stability diagram for the stripe pattern
analogous to the Busse balloon \cite{busse} for Rayleigh-B\'enard
convection. We take the maximum value of $\left<k\right>$ before a
stripe pinches off as the wave number corresponding to the
skew-varicose instability, and the minimum value of $\left<k\right>$
before the appearance of crossrolls as that for the crossroll
instability.  

\begin{figure*}[htbp]
\epsfxsize=3in
\centerline{\epsffile{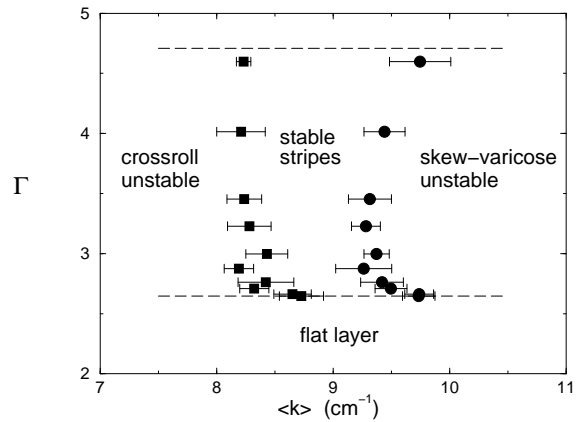}}
\vskip 0.1in
\caption{The stability boundary for stripe patterns at $N = 9.2$, $f =
29.7$ Hz. Squares indicate the crossroll instability and circles the
skew-varicose instability. Each point is the average over several
events; the error bars are standard deviations. The lower dashed line
indicates the point at which the pattern fills the cell as $\Gamma$ is
increased, and the upper dashed line marks the transition from stripes
to hexagons.}
\label{stability}
\end{figure*}

Figure \ref{stability} is the resulting stability diagram for $N=
9.2$ and $f=29.7$ Hz. It indicates that there is a limited range of
$k$ over which the stripe pattern is stable. The stability range is
bounded at high $k$ by the skew-varicose instability and at low $k$ by
the crossroll instability.  Similar stability boundaries, but with
slightly different shapes, were measured at a number of different
values of $N$ and $f$. However, for the depth in Fig. \ref{stability},
the boundary is different at high frequencies: At $f = 40$ Hz, the
stripes are unstable to crossrolls on both the high and low $k$ sides
of the stability boundary, and at the same depth and $ f = 45$ Hz,
squares rather than hexagons limit the range of existence of stripes
at high $\Gamma$. These stability boundaries are very similar to those
for Rayleigh-B\'enard convection rolls: For low Prandtl number fluids
\cite{prandtl}, the skew-varicose instability forms the high-$k$ limit
of the stability balloon, while crossrolls are observed on the low-$k$
side \cite{busse,pcl85,svrbc}. For higher Prandtl number, the
crossroll instability occurs at both high and low $k$
\cite{busse,crrbc}.

\begin{figure*}[htbp]
\epsfxsize=3.375in
\centerline{\epsffile{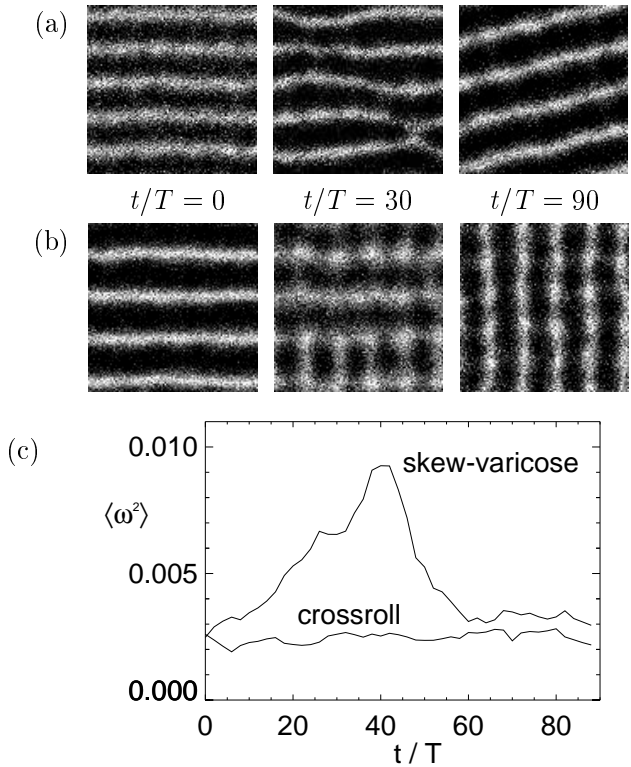}}
\vskip 0.1in
\caption{Instabilities of a stripe pattern from simulations of the
oscillating layer with $\Gamma = 3.0$, $N=5.4$. (a) Skew-varicose
instability at $f/\sqrt{g/H} = 0.37$; $kH=1.7$ initially and and 1.4
after the instability. (b) Crossroll instability at $f/\sqrt{g/H} =
0.40$; $kH= 1.36$ before and 1.7 after. Here $H$ is the layer depth
and $T$ is the oscillation period. (c) Mean-square vertical vorticity
as a function of time for the instabilities shown in (a) and (b).}
\label{simfig}
\end{figure*}

We have also observed skew-varicose and crossroll instabilities in
event-driven molecular dynamics simulations which have been previously
validated against experiment \cite{bssms98}. We simulated the motions
of 60000 particles in a square cell of side 100 particle diameters,
with periodic boundary conditions in both horizontal directions. A
pattern of straight stripes with $k$ either too large or too small to
be stable at the frequency of the simulation was used as the initial
condition. Time sequences of the resulting instabilities are shown in
Fig. \ref{simfig}. As in the experiments, the skew-varicose
instability leads to a decrease in $k$ while the crossroll instability
leads to an increase in $k$.

Patterns of rolls in fluid convection are well described by partial
differential equations for the amplitude (near onset) and the phase of
the pattern \cite{ch93}. The crossroll instability exists in an
amplitude equation description that allows for the existence of two
perpendicular modes. The skew-varicose instability, however, depends
on the existence of a large scale vertical vorticity (or mean flow),
which couples to roll curvature \cite{c83}.  To determine whether an
analogous vertical vorticity exists for skew-varicose instabilities in
the granular layer, the particle velocities from the simulations were
averaged spatially over depth and temporally over one oscillation of
the pattern. The resulting two-dimensional velocity field $\vec v$ was
then low-pass filtered to remove spatial frequencies due to the
pattern itself.  Finally, we calculated the squared vorticity,
$\omega^2 \equiv (\nabla \times \vec v)^2$, and averaged it over the
entire cell. The result is shown in Fig. \ref{simfig}(c).  The
vertical vorticity increases as the skew-varicose instability
develops, then sharply decreases after a stripe pinches off.  In
contrast, there is no change in $\left<\omega^2\right>$ for the
crossroll instability.

We have shown that the range of stability of stripe patterns which
form in an oscillated granular layer is limited by the skew-varicose
and crossroll instabilities, and that the stability boundary of stripe
patterns in granular layers is similar to that of straight rolls in a
convecting fluid. Our simulations produce the same instabilities, and
indicate that vertical vorticity plays a role in the skew-varicose
instability, as it does in fluid systems.  These results demonstrate a
clear correspondence between patterns in oscillated granular media and
hydrodynamic systems, and suggest that, despite the unique physical
properties of the oscillated granular layer, a continuum description
of this system should be possible.

We are grateful to W. D. McCormick and B. B. Plapp for helpful
discussions.  This research was supported by the Department of Energy
Office of Basic Energy Sciences and the Texas Advanced Research
Program.


\begin{references}

\bibitem[*]{perm} Permanent address: Department of
Physics and Physical Oceanography, Memorial University of
Newfoundland, St. John's, Newfoundland, Canada A1B 3X7. Email address:
jdebruyn@kelvin.physics.mun.ca
\bibitem{jnb96} H. M. Jaeger, S. R. Nagel, and R. B. Behringer,
Rev. Mod. Phys. {\bf 68}, 1259 (1996); L. P. Kadanoff, unpublished.
\bibitem{dlk95} Y. Du, H. Li, and L. P. Kadanoff, Phys. Rev. Lett. 
{\bf 74}, 1268 (1995). 
\bibitem{haff83} P. K. Haff, J. Fluid Mech. {\bf 134}, 401 (1983).
\bibitem{mus95} F. Melo, P. B. Umbanhowar, and H. L. Swinney, Phys. Rev. Lett.
{\bf 75}, 3838 (1995).
\bibitem{ums97} P. B. Umbanhowar, F. Melo, and H. L. Swinney, Physica
A, in press.
\bibitem{ums96} P. B. Umbanhowar, F. Melo, and H. L. Swinney, Nature (U. K.)
{\bf 382}, 793 (1996).
\bibitem{ch93} M. C. Cross and P. C. Hohenberg, Rev. Mod. Phys. {\bf 65}, 851 
(1993).
\bibitem{busse} F. H. Busse, Rep. Prog. Phys. {\bf 41}, 1929 (1978);
F. H. Busse, in {\it Hydrodynamic Instabilities and the Transition to
Turbulence}, edited by H.L. Swinney and J.P. Gollub (Springer, Berlin,
1981), p. 97.
\bibitem{kg96} A. Kudrolli and J. P. Gollub, Physica {\bf D97}, 133 (1996).
\bibitem{lcrd96} S. Luding, E. Cl\'ement, J. Rajchenbach, and
J. Duran, Europhys. Lett. {\bf 36}, 247 (1996); L. Labous and
E. Cl\'ement, unpublished; K. M. Aoki and T. Akiyama,
Phys. Rev. Lett. {\bf 77}, 4166 (1996); C. Bizon {\it et al.}, {\it
ibid.} {\bf 79}, 4713, (1997); K. M. Aoki and T. Akiyama, {\it ibid.}
{\bf 79}, 4714 (1997).
\bibitem{bssms98} C. Bizon, M. D. Shattuck, J. B. Swift, W. D. 
McCormick, and H. L. Swinney, Phys. Rev. Lett. {\bf 80}, 57 (1998).
\bibitem{shin97} T. Shinbrot, Nature (U. K.) {\bf 389}, 574 (1997);
E. Cerda, F. Melo, and S. Rica, Phys. Rev. Lett. {\bf 79}, 4570
(1997); D. H. Rothman, unpublished; J. Eggers and H. Riecke,
unpublished.
\bibitem{ta97} L. S. Tsimring and I. S. Aranson, Phys. Rev. Lett. {\bf 79}, 213 (1997);
H. Sakaguchi and H. R. Brand, J. Phys. II, 
{\bf 7}, 1325 (1997).
\bibitem{vo97} S. C. Venkataramani and E. Ott, unpublished.
\bibitem{pcl85} A. Pocheau, V. Croquette, and P. LeGal, Phys. Rev. Lett. 
{\bf 55}, 1094 (1985); B. B. Plapp, Ph. D. thesis, Cornell University, 1997
(unpublished).
\bibitem{svrbc}  Y. Hu, R. Ecke, and G. Ahlers, Phys. Rev. E
{\bf 48}, 4399 (1993).
\bibitem{ef94} W. S. Edwards and S. Fauve, J. Fluid Mech. 
{\bf 278}, 123 (1994).
\bibitem{crrbc}
F. H. Busse and J. A. Whitehead, J. Fluid Mech. {\bf 47}, 305 (1971).
\bibitem{emb97} D. Egolf, I. Melnikov, and E. Bodenschatz, to be
published.  This is an efficient way to determine $\vec k$ for
stripe-like patterns, but does not give correct results for square
patterns or at defects.
\bibitem{prandtl} The Prandtl number of a fluid is the ratio of its
viscosity to its thermal diffusivity.
\bibitem{c83} M. C. Cross, Phys. Rev. A {\bf 27}, 490 (1983).

\end{references}
\end{document}